\PassOptionsToPackage{unicode}{hyperref}
\PassOptionsToPackage{hyphens}{url}
\documentclass[
  11pt,
]{article}
\usepackage{lmodern}
\usepackage{amssymb,amsmath}
\usepackage{ifxetex,ifluatex}
\ifnum 0\ifxetex 1\fi\ifluatex 1\fi=0 
  \usepackage[T1]{fontenc}
  \usepackage[utf8]{inputenc}
  \usepackage{textcomp} 
\else 
  \usepackage{unicode-math}
  \defaultfontfeatures{Scale=MatchLowercase}
  \defaultfontfeatures[\rmfamily]{Ligatures=TeX,Scale=1}
\fi
\IfFileExists{upquote.sty}{\usepackage{upquote}}{}
\IfFileExists{microtype.sty}{
  \usepackage[]{microtype}
  \UseMicrotypeSet[protrusion]{basicmath} 
}{}
\makeatletter
\@ifundefined{KOMAClassName}{
  \IfFileExists{parskip.sty}{%
    \usepackage{parskip}
  }{
    \setlength{\parindent}{0pt}
    \setlength{\parskip}{6pt plus 2pt minus 1pt}}
}{
  \KOMAoptions{parskip=half}}
\makeatother
\usepackage{xcolor}
\IfFileExists{xurl.sty}{\usepackage{xurl}}{} 
\IfFileExists{bookmark.sty}{\usepackage{bookmark}}{\usepackage{hyperref}}
\hypersetup{
  pdftitle={Untitled},
  hidelinks,
  pdfcreator={LaTeX via pandoc}}
\usepackage[margin=1in]{geometry}
\usepackage{longtable,booktabs}
\usepackage{etoolbox}
\makeatletter
\patchcmd\longtable{\par}{\if@noskipsec\mbox{}\fi\par}{}{}
\makeatother
\IfFileExists{footnotehyper.sty}{\usepackage{footnotehyper}}{\usepackage{footnote}}
\makesavenoteenv{longtable}
\usepackage{graphicx,grffile}
\makeatletter
\def\maxwidth{\ifdim\Gin@nat@width>\linewidth\linewidth\else\Gin@nat@width\fi}
\def\maxheight{\ifdim\Gin@nat@height>\textheight\textheight\else\Gin@nat@height\fi}
\makeatother
\setkeys{Gin}{width=\maxwidth,height=\maxheight,keepaspectratio}
\makeatletter
\def\fps@figure{htbp}
\makeatother
\usepackage{booktabs}
\usepackage{longtable}
\usepackage{array}
\usepackage{multirow}
\usepackage{spreadtab}
\usepackage{wrapfig}
\usepackage{float}
\usepackage{colortbl}
\usepackage{pdflscape}
\usepackage{tabu}
\usepackage{threeparttable}
\usepackage{threeparttablex}
\usepackage[normalem]{ulem}
\usepackage{makecell}
\usepackage{xcolor}
\usepackage{todonotes}
\usepackage[]{natbib}
\usepackage[export]{adjustbox}
\usepackage{color,soul}
\usepackage{enumerate}
\usepackage{siunitx}
\usepackage{authblk}
\usepackage{caption} 

\usepackage{authblk}

\usepackage{authblk}

\title{The Extended Crosswise Model Adjusted for Random Answering}
\author[1,2]{Khadiga H. A. Sayed \footnote{Corresponding author: k.h.a.sayed@uu.nl}}
\author[1]{Maarten J. L. F. Cruyff}
\author[3,4]{Andrea Petr{\'o}czi}
\author[1,5]{Peter G. M. van der Heijden}

\affil[1]{Department of Methodology and Statistics, Utrecht University, Utrecht, The Netherlands}
\affil[2]{Department of Statistics, Faculty of Economics and Political Science, Cairo University, Egypt}
\affil[3]{Kingston University London, School of Life Sciences, Pharmacy and Chemistry, Faculty of Health, Science, Social Care and Education, London, UK}
\affil[4]{ELTE E{\"o}tv{\"o}s Lor{\'a}nd University, Institute of Health Promotion and Sport Sciences, Faculty of Education and Psychology, Budapest, Hungary}
\affil[5]{Department of Social Statistics and Demography, University of Southampton, Southampton, UK}

\date{\vspace{-2.5em}}    

\begin{document}
\maketitle

\hypertarget{abstract}{%
\section*{Abstract}\label{abstract}}

The Extended Crosswise Model is a popular randomized response design that employs a sensitive  and a randomized innocuous statement, and asks respondents if one of these statements is true, or that none or both are true. The model has a degree of freedom to test for response biases, but is unable to detect random answering. In this paper, we propose two new methods to indirectly estimate and correct for random answering. One method uses a non-sensitive control statement and a quasi-randomized innocuous statement to which both answers are known to estimate the proportion of random respondents. The other method assigns less weight in the estimation procedure to respondents who complete the survey in an unrealistically short time. For four surveys among elite athletes, we use these methods to correct the prevalence estimates of doping use for random answering.          

\textbf{Keywords:} Randomized response, innocuous statement, sensitive attribute,  goodness-of-fit test,  number sequence randomizer, doping use

\hypertarget{introduction}{%
\section{Introduction}\label{introduction}}

Randomized response technique is an interview method introduced by \cite{warner1965} to eliminate evasive response biases when sensitive questions have to be asked. It involves the use of  a randomizer (e.g., a die or a spinner) that perturbs respondents' answers so that no direct link can be established with their true, non-randomized answers. It has been shown that Warner's design and many of its variants do not completely eliminate evasive response bias due the respondents' reluctance to (falsely) incriminate themselves \citep{boeije2002}. For this reason randomized response designs have been proposed that avoid the use of incriminating responses. The most well known are the crosswise model (CWM, \cite{yu2008}) and its extension, the extended CWM (ECWM, \cite{heck2018}). In these designs, respondents are shown two statements; one about a sensitive attribute with unknown prevalence, e.g., ``I have used illegal drugs'', and an innocuous one with known prevalence, e.g., ``I was born in January or February''. Respondents are instructed to indicate whether their answer is ``DIFFERENT'', i.e. one `yes' and one `no' answer, or ``SAME'', i.e. two `yes' or two `no' answers. Unlike designs such as the unrelated question model \citep{greenberg} and the triangular design \citep{meisters2022new, hsieh2024prevalence}, these neutral answer options make it easier to give an honest answer, while at the same time they make it more difficult to infer the incriminating answer, and thus to give a evasive answer. The statistical model of the CWM is saturated, i.e. it has only one non-redundant randomized response proportion to estimate the prevalence of the sensitive attribute, and therefore does not allow for a goodness-of-fit test. The ECWM extends the CWM by randomly splitting the sample into two non-overlapping sub-samples with  complementary  probabilities of answering `yes' to the innocuous statement. The model for this design has a degree of freedom that allows for a goodness-of-fit test.

The (E)CWM has been investigated in several validation studies of socially undesirable attributes such as plagiarism \citep{coutts2011,jann2012,hopp2019}, tax evasion \citep{korndorfer2014}, xenophobia and Islamophobia \citep{hoffmann2016, hoffmann2020validity, meisters2020controlling}, socially desirable attributes such as personal hygiene behaviour during the COVID-19 pandemic \citep{mieth2021}, and voluntary work in the social sector \citep{meisters2022more}. Compared to the prevalence estimates obtained with the direct questioning method, the prevalence estimates of the (E)CWM were higher for undesirable attributes and lower for desirable ones, and therefore considered more valid according to the  ``more/less-is-better'' criterion \citep{umesh1991,sagoe2021, schnell2021}. However, the (E)CWM is considered be more prone to random answering than other randomized response designs. The reason for this is that the meaning of its answer categories ``DIFFERENT'' and ``SAME'' is not well understood by the respondents, and may therefore incite indifference. In studies where respondents were asked directly if they had answered the (E)CWM statement randomly, 2 to 19\% of the respondents admitted having answered randomly \citep{enzmann2017, schnapp2019, meisters2020}.

Some studies suggest that the higher presumed validity of the (E)CWM prevalence estimates according to the ``more/less is better criterion'' may (partly) be explained by random answering \citep{hoglinger2016, hoglinger2018, enzmann2017, hoglinger2017}, as random answering biases the prevalence estimates towards 50\% \citep{walzenbach2019}. Therefore it is important to detect and correct for random answering. The degree of freedom of the ECWM can detect systematic response biases such as preferring one answer option as being more safe \citep{heck2018, cruyff2024one}, but it is unable to detect random answering \citep{heck2018, meisters2022more}. As a consequence, other solutions have been proposed to address the issue of random answering. One is to provide detailed instructions, and using comprehension checks to detect and remove random responders from the data \citep{hoglinger2017, hoglinger2018, meisters2020}. This method presupposes that the sensitive attribute is not associated with random answering.  A study by \citet{meisters2022more} investigated how the higher validity of the (E)CWM is affected by random answering behaviour through manipulating two factors: the direction of social desirability (undesirable vs. desirable) and the prevalence of sensitive attributes (high vs. low). This study found a minor effect of random answering on the estimates according to the  ``more/less-is-better'' criterion. \cite{enzmann2017} and \cite{schnapp2019} suggested estimating the prevalence of random responders by asking the respondents whether they had answered the statements at random, and adjusting the (E)CWM estimate accordingly. This approach assumes that random responders answer this (sensitive) question honestly instead of randomly. Alternatively, \cite{atsusaka2021} suggested estimating the prevalence of random responders using a sensitive anchor statement with known prevalence. For instance, for sensitive statement of interest is ``In order to avoid paying a traffic ticket, I would be willing to pay a bribe to a police officer'', the anchor statement is ``I have paid a bribe to be on the top of a waiting list for an organ transplant'', with a known prevalence of zero. The challenge of this method is to formulate a relevant sensitive anchor statement that matches the sensitive statement of interest. 

\begin{figure}[!ht]
    \centering
    \includegraphics[width=.8\textwidth]{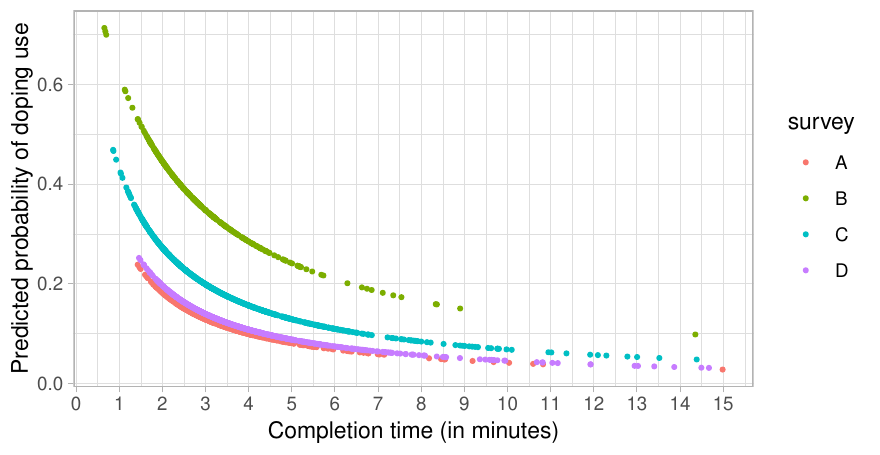}
    \caption{The predicted probabilities of doping use as function in survey completion time.}
    \label{fig:doping}
\end{figure}

\begin{figure}[!ht]
\centering
\includegraphics[width=1\textwidth]{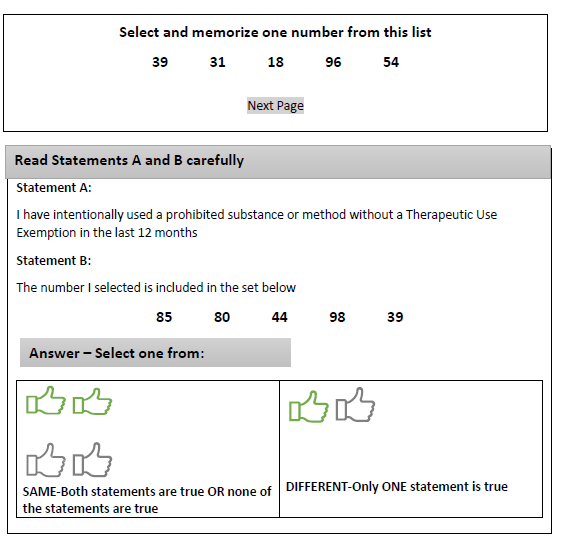}
\caption{Example of the number sequence randomizer for the statement on doping use.}\label{fig:example}
\end{figure}

In this paper, we propose two new methods to estimate and correct for random answering in the (E)CWM design. The first method involves a non-sensitive control statement with known prevalence of 0 or 1 in combination with an innocuous statement with the probability of a `yes' answer set to either 0 or 1. To mimic the ordinary ECWM procedure as closely as possible, the suggestion is raised that the answers to the innocuous question are randomized. This is achieved by using the number sequence randomizer \citep{sayed2022refinement}, which asks respondents to memorize one number from a sequence of five numbers and to indicate whether this number reappears in a second sequence of five, in which either all or none of the numbers from the first sequence reappear (for an example, see Figure \ref{fig:example} in Section~\ref{data}).  This allows us to check whether the answer of each individual respondent is correct or not. By attributing the proportion of incorrect ``DIFFERENT'' or ``SAME'' answers to random answering, an estimate of the total proportion of random answers is obtained.  The second method uses a function of the respondents' time to complete the survey as a weight in the log-likelihood function of the ECWM, so that potential random respondents exert less influence on the prevalence estimate than attentive respondents. The idea behind this method is that fast respondents are more likely to answer randomly than respondents who took more time to answer the questions, and thus should be assigned a smaller weight. The logistic regression analyses of the data of Surveys  A to D shown in Figure \ref{fig:doping} seem to corroborate this conjecture. Random answering biases the prevalence estimates of doping use toward 50\%, and the predicted probabilities of the fast respondents are closer to 50\% than those of the slower ones.

To illustrate the two new methods, they are applied to data from four ECWM surveys on doping use by elite athletes. These data were analyzed before by \cite{cruyff2024one}, who found convincing evidence for  \textit{one-saying}. The term one-saying is derived from the answer option ``I have One `yes' and One `no' answer'' used in some of surveys instead of the equivalent answer option ``DIFFERENT''. In order to take one-saying into account, we develop our methods for both the standard ECWM as for the one-sayers model.  

The paper is structured as follows. Section~\ref{data} provides a description of the data, consisting of four surveys on doping use among elite athletes. Section~\ref{model} presents reviews of the ECWM and the one-sayers model, and extends these models to account for random answering. Section~\ref{correction} presents the two methods to account for random answering . Section~\ref{results} presents the prevalence estimates of doping corrected for random answering. Section~\ref{discuss}  discusses the results of our analyses and ends with some concluding remarks.

\hypertarget{data}{%
\section{Data}\label{data}}

The data are from four surveys on doping use which were conducted as a part of the World Anti-Doping Agency (WADA) anti-doping program. A previous analysis of these data showed the presence of one-saying \citep{cruyff2024one}. Data analysis was approved by the Ethics Review Board of the Faculty of Social and Behavioural Sciences of Utrecht University in application 22-0185.

The respondents in these surveys were elite athletes over the age of 16 years. The numbers of athletes that completed the surveys A to D were respectively 354, 325, 915, and 813. In these surveys the sensitive statement ``I have intentionally used a prohibited substance or method without a Therapeutic Use Exemption (TUE) in the last 12 months.'' was paired with the number sequence randomizer as the innocuous statement \citep{sayed2022refinement}. Figure~\ref{fig:example} shows how this works. Respondents are first presented with a sequence of five randomly generated two-digit numbers, and are asked to memorize one. Then they are shown a second sequence of five numbers in which either one or four numbers of the first sequence reappear, with the innocuous statement B asking whether the memorized number is in this sequence. In Figure 2 only one number reappears. The respondents are then asked to indicate whether their answers to these two statements are the ``SAME'' (both statements are true or false) or ``DIFFERENT'' (only one statement is true). In this example, the probability of answering `yes' to the innocuous statement B is $p=1/5$ because only one of the five numbers from the first sequence reappears in the second. In the surveys, respondents were randomly assigned to a condition with $p=1/5$ (when one number reappears) or with $p=4/5$ (when four numbers reappear).

To check the understanding and adherence to the ECWM instructions, the athletes were also presented with the control question A ``I am a licensed/accredited athlete'' in combination with a number sequence statement B. The answers to statement B were quasi-randomized, because the probability of a `yes' answer set either to 1 by letting all the numbers of the first sequence reappear in the second sequence, or to 0 by letting none of the numbers reappear. The probability of a `yes' answer to the control question was assumed to be 1, because for all athletes in the surveys having a license/accreditation is mandatory. However, it appeared that the athletes were not always aware of being licensed/accredited. The reason for this is that the license/accreditation it is often arranged by their sports associations. Consequently, some of the athletes may have believed that they were not licensed/accredited, and may therefore have given the wrong answer to the control question. 

Surveys A and D were accessed via a web-link and administered on an online platform, while Surveys B and C also offered the possibility of administration on mobile phones/tablets. The time to complete the survey was recorded when the respondents clicked a button ``End survey'' that stopped the timer. The completion time were not measured without error, because it was sometimes forgotten to click the ``Submit'' button  resulting in unrealistically high completion times. 

\hypertarget{model}{%
\section{The models}\label{model}}

This section reviews the ECWM and the one-sayers model using matrix notation, and extends both models to account for random answering.

\hypertarget{ecwm}{%
\subsection{The ECWM}\label{ecwm}}

Consider a CWM with a sensitive  statement with unknown prevalence $\pi$ and an innocuous statement with known randomization probability $p$ of answering it with `yes' (e.g. ``Is your birthday in the first two months of the year?'').  Let $\pi^*_{y}$ be the probability of observing randomized response $y$, for $y\in\{1\equiv\mbox{``DIFFERENT''}, ~2\equiv\mbox{``SAME''}\}$, $\pi$ the prevalence of the sensitive attribute, and $p$ the randomization probability of answering the innocuous statement with `yes', for $p=1-q\neq .5$. In matrix notation the model consists of the vectors $\boldsymbol\pi^*$ with the randomized response probabilities and $\boldsymbol\pi$ with the probabilities that the sensitive attribute is present or absent, and a $2\times2$ transition matrix with the randomization probabilities:

\begin{eqnarray} \left(\label{mod:cwm}
\begin{array}{c}
\pi^*_1\\
\pi^*_2
\end{array}\right) =
\left(\begin{array}{ccc}
p & q\\
q & p
\end{array}\right)
\left(\begin{array}{c}
\pi\\
1-\pi
\end{array}\right)=
\left(\begin{array}{ccc}
p\pi + q(1-\pi)\\
q\pi + p(1-\pi)
\end{array}\right).
\end{eqnarray}

The ECWM divides the sample in two sub-samples with the respective complementary probabilities $p$ and $q$ of answering `yes' to the innocuous statement. Let $\pi^*_{y\mid s}$ be the conditional probability of observing randomized response $y$ given membership of sub-sample $s$, for $s\in\{1,2\}$. In matrix notation, the ECWM is given by

\begin{eqnarray} \left(\label{mod:adherence}
\begin{array}{c}
\pi^*_{1|1}\\
\pi^*_{2|1}\\
\pi^*_{1|2}\\
\pi^*_{2|2}
\end{array}\right) =
\left(\begin{array}{ccc}
p & q\\
q & p\\
q & p\\
p & q
\end{array}\right)
\left(\begin{array}{c}
\pi\\
1-\pi
\end{array}\right)=
\left(\begin{array}{ccc}
p\pi + q(1-\pi)\\
q\pi + p(1-\pi)\\
q\pi + p(1-\pi)\\
p\pi + q(1-\pi)
\end{array}\right).
\end{eqnarray}

Eq. (\ref{mod:adherence}) shows that $\pi^*_{1|1}=\pi^*_{2|2}$ and $\pi^*_{2|1}=\pi^*_{1|2}$. The model estimates the parameter $\pi$ using two  non-redundant observed response frequencies (as the conditional probabilities $\pi^*_{y\mid s}$ within each sub-sample $s$ need to sum to 1) and therefore has one degree of freedom. 

\hypertarget{random answer}{%
\subsection{The ECWM correcting for random answering}\label{ecwm-1}}

To account for random responders in the ECWM, let $\gamma$ denote the prevalence of random responders. Assuming equal probabilities of random answering for all randomized responses, the model is given by

\begin{eqnarray} \left(\label{mod:random}
\begin{array}{c}
\pi^*_{1|1}\\
\pi^*_{2|1}\\
\pi^*_{1|2}\\
\pi^*_{2|2}
\end{array}\right)=
\left(\begin{array}{ccc}
(1-\gamma)p +  .5\,\gamma &   (1-\gamma)q +  .5\,\gamma\\
(1-\gamma)q +  .5\,\gamma &   (1-\gamma)p +  .5\,\gamma\\
(1-\gamma)q +  .5\,\gamma &   (1-\gamma)p +  .5\,\gamma\\
(1-\gamma)p +  .5\,\gamma &  (1-\gamma)q +  .5\,\gamma
\end{array}\right)
\left(\begin{array}{c}
\pi\\
1-\pi
\end{array}\right),
\end{eqnarray}
We  refer to this model as ``ECWM+RA'' where RA stands for random answer. In Eq. (\ref{mod:random}) the equality relations $\pi^*_{1|1}=\pi^*_{2|2}$ and $\pi^*_{2|1}=\pi^*_{1|2}$ are not affected by random answering. This means that multiple combinations of $\gamma$ and $\pi$ give the same parameter vector $\boldsymbol\pi^*$, and therefore the model is not identified, as was noted before by \cite{heck2018}.

\hypertarget{onesayer}{%
\subsection{The one-sayers  model correcting for random answering}\label{one-sayer model}}

The one-sayers model \citep{cruyff2024one} accounts for evasive respondents who answer ``DIFFERENT'' (or equivalently ``I have One ‘yes‘ and One ‘no’ answer''), irrespective of the outcome of the randomizer. With $\theta$ denoting the prevalence of one-sayers, the model is given by

\begin{eqnarray} \left(\label{mod:one-saying}
\begin{array}{c}
\pi^*_{1|1}\\
\pi^*_{2|1}\\
\pi^*_{1|2}\\
\pi^*_{2|2}
\end{array}\right) =
\left(\begin{array}{ll}
(1 - \theta)p + \theta  & (1 - \theta)q + \theta\\
(1 - \theta)q  & (1 - \theta)p \\
(1 - \theta)q + \theta  & (1 - \theta)p + \theta  \\
(1 - \theta)p & (1 - \theta)q 
\end{array}\right)
\left(\begin{array}{c}
\pi\\
1-\pi
\end{array}\right),
\end{eqnarray}
This model shows that $\pi^*_{1|1}\neq\pi^*_{2|2}$ and $\pi^*_{2|1}\neq\pi^*_{1|2}$ for $\theta>0$.  Now each combination of $\theta$ and $\pi$ yields a unique parameter vector $\boldsymbol\pi^*$, and therefore the model is identified. 

The one-sayers model that additionally accounts for random answering (One-sayers + RA) is given by
\begin{eqnarray} \left(\label{mod:onesayer-rr}
\begin{array}{c}
\pi^*_{1|1}\\
\pi^*_{2|1}\\
\pi^*_{1|2}\\
\pi^*_{2|2}
\end{array}\right)=
\left(\begin{array}{ll}
(1-\gamma - \theta)p +  \theta + .5\,\gamma &   (1-\gamma - \theta)q + \theta+ .5\,\gamma\\
(1-\gamma-\theta)q +  .5\,\gamma &   (1-\gamma-\theta)p +  .5\,\gamma\\
(1-\gamma-\theta)q + \theta+ .5\,\gamma &   (1-\gamma-\theta)p + \theta+ .5\,\gamma\\
(1-\gamma-\theta)p +  .5\,\gamma &  (1-\gamma-\theta)q +  .5\,\gamma
\end{array}\right)
\left(\begin{array}{c}
\pi\\
1-\pi\
\end{array}\right),
\end{eqnarray}
with the restriction that $\theta + \gamma \leq1$ because one-saying and random answering are mutually exclusive response categories. This model is over-parameterized, but it can be identified  by fixing the parameter $\gamma$. To do so, a reasonable estimate of the prevalence of random responders has to obtained by other means (e.g. by the use of a control question).

\subsection{Expected bias resulting from non-zero $\gamma$ and $\theta$ parameters}

\begin{figure}[!ht]
    \centering
    \includegraphics[width=.9\textwidth]{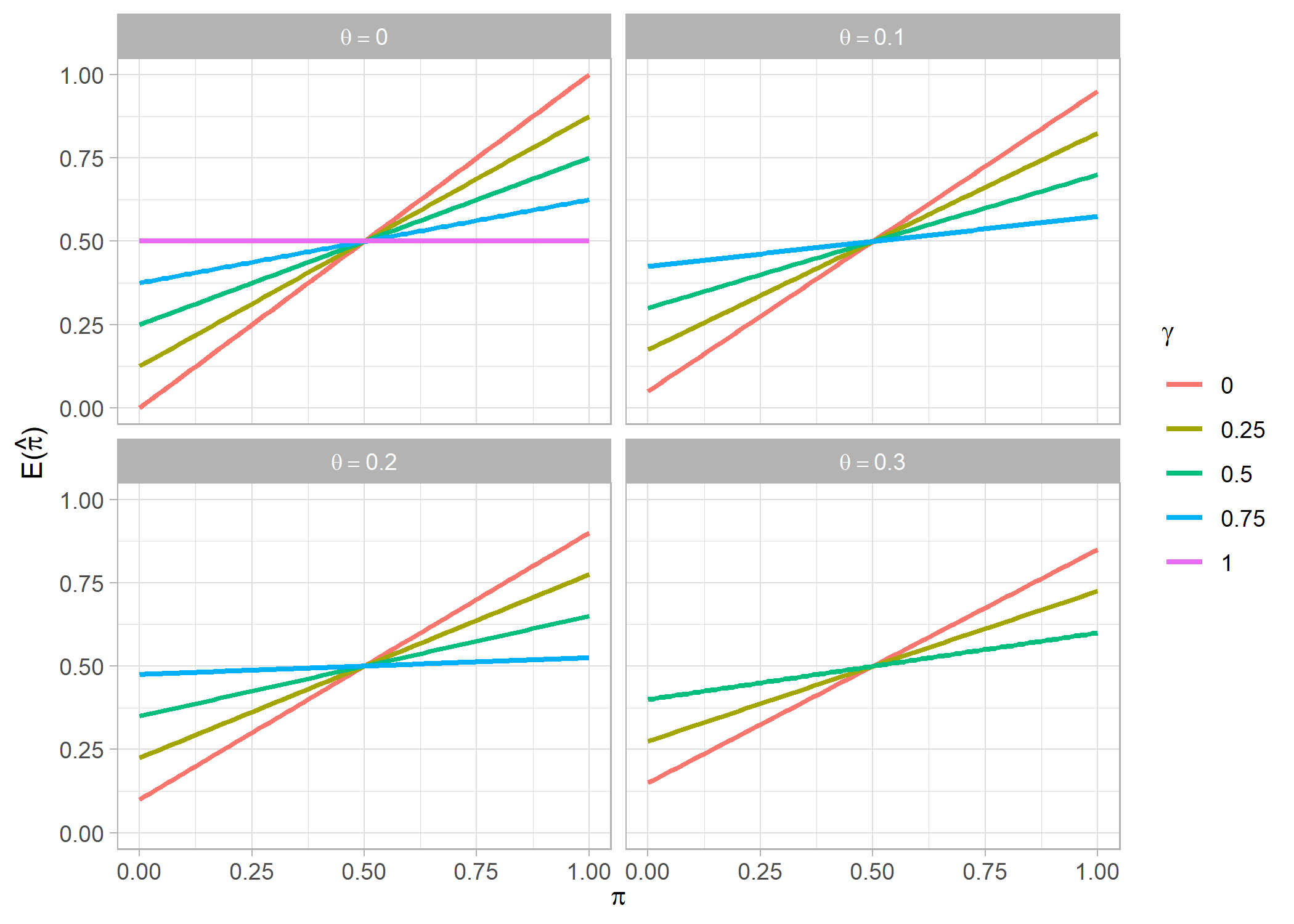}
    \caption{The expectation $\mathbb{E}(\hat\pi)$ under the ECWM as a function $\gamma$, $\theta$ and $\pi$.}
    \label{fig:bias}
\end{figure}

In the presence of one-saying and random responding, the standard ECWM estimator $\hat\pi$ of model (\ref{mod:adherence}) yields a biased estimate of $\pi$ (i.e., $\mathbb{E}(\hat\pi) = \pi + Bias$). The size of this bias equals $(\theta + \gamma) (.5 -\pi)$ (see Appendix A on OSF (\url{https://osf.io/ekcjb/?view_only=e5ad20e51f2c4b4ea3816d5281350782}) for a derivation). Figure \ref{fig:bias} shows the expected estimates of $\pi$ when response bias due to random answering and one-saying is not taken into account. 

The top-left plot in Figure \ref{fig:bias} depicts the expected estimates of $\pi$ when random answering is present but there is no one-saying. It shows that as the value of $\gamma$ increases, the expected estimate of $\pi$ approaches 50\%. In the case that all respondents answer randomly, $\mathbb{E}(\hat\pi)=0.5$, irrespective of the value of $\pi$. The remaining three plots show the expected estimates of $\pi$ for the prevalence of one-saying $\theta\in\{.1, .2, .3\}$ with the restriction that $\gamma + \theta\leq 1$, because one-saying and random answering are two mutually exclusive phenomenon. This explains why less values of $\gamma$ are shown with the increasing values of $\theta$. The plots show similar patterns for $\gamma$ as in the top-left panel, but with higher expected estimates of $\pi$  as the value of $\gamma$ increases. For instance when $\pi =.25$, the standard ECWM of (\ref{mod:adherence}) yields biased expected estimates of  $\hat\pi\in\{.31, .38, .44, .5\}$ for $\gamma\in\{.25, .5, .75, 1\}$, and $\theta = 0$. For $\gamma\in\{.25, .5, .75\}$, and $\theta =.1$, the ECWM  yields biased expected estimates of $\hat\pi\in\{.34, .4, .46\}$, respectively. In summary, the figure shows that both random answering and one-saying bias the prevalence estimates towards 0.5.

\subsection{Estimation}

This section presents moment and maximum likelihood estimates of the model parameters presented in the previous section.

\subsubsection*{Moment estimation}

The unidentified ECWM  correcting for random answering  (\ref{mod:random}) is identified by using a value for $\gamma$ derived from the non-sensitive control statement. The  estimator $\hat\pi_{ra}$ of the sensitive attribute is given by

\begin{equation}\label{eq:pi-corrected}
    \hat\pi_{ra}= \frac{\pi^*_{11} + \pi^*_{22} - (1-\gamma)q - .5 \gamma}{(p-q) (1-\gamma)}.
\end{equation} 

where $\pi^*_{ys} =(n_s/n) \pi^*_{y\mid s}$ is the unconditional probability of observing response $y$ in sub-sample $s$, $n_s$ is the sub-sample size and $n=\sum_sn_s$ the total sample size. The moment estimator in Eq. (\ref{eq:pi-corrected}) is computed by plugging in the  observed sample proportions $n_{ys}/n$  as estimates for $\pi^*_{ys}$. 

This estimator is identical to the ones presented by \cite{schnapp2019} and \cite{atsusaka2021}, but the difference is in the method of estimating $\gamma$. \cite{schnapp2019} estimates $\gamma$ by directly asking respondents whether or not they answered the crosswise statement randomly. Asking a direct question ensures that $\gamma$ has minimal (binomial) variance, but because it is sensitive makes it is prone to evasive response bias. \cite{atsusaka2021} avoid this bias by using the crosswise design for the sensitive anchor statement, but this results in a larger (randomized response) variance. Our estimate of $\gamma$ is derived from a non-sensitive control statement with know prevalence of 0 or 1. Since this statement is not sensitive, the answers are not affected by evasive response bias. This statement is combined with a number sequence statement with probability of answering `yes' set to either 0 or 1 eliminating the randomization of the responses, which makes it statistically equivalent to a direct question. As a consequence, ``DIFFERENT'' and ``SAME'' are also answers to a direct question, and therefore our estimate of $\gamma$ also has minimal (binomial) variance.

For the model one-sayers model correcting for random answering (\ref{mod:onesayer-rr}), the moment estimator of $\theta$ is

\begin{equation}\label{eq:theta-RR}
    \hat\theta = \pi^*_{1\mid1}\,+\, \pi^*_{1\mid2}  -1,
\end{equation}

which is identical to the estimator of one-sayers in model (\ref{mod:one-saying}), i.e., the presence of random responders does not affect the  estimate of one-sayers. The estimator $\hat\pi_{one+ra}$ of the sensitive attribute is

\begin{equation}\label{eq:pihat-RR-one}
 \hat\pi_{one+ra} =  \frac{p\pi^*_{2\mid 2}-q\pi^*_{2\mid 1} -\gamma (p-.5)}{
 (p-q)\, (\pi^*_{2\mid 1} + \pi^*_{2\mid 2} -\gamma)},\quad p\neq 0.5 
\end{equation}

The analytical variances of the estimators $\hat\pi_{ra}$ and $\hat\pi_{ra+one}$ are presented in Appendix A on OSF (\url{https://osf.io/ekcjb/?view_only=e5ad20e51f2c4b4ea3816d5281350782}). Substituting $\gamma = 0$ in Eqs. (\ref{eq:pi-corrected}) and (\ref{eq:pihat-RR-one}) yields the moment estimators of $\pi$ of the standard ECWM of (\ref{mod:adherence}), and the one-sayers model of (\ref{mod:one-saying}), respectively.  

\subsubsection*{Maximum likelihood estimation}
The parameters $\pi$ and/or $\theta$ of models (\ref{mod:random}) and (\ref{mod:onesayer-rr})  can alternatively be estimated by maximization of the log-likelihood 
\begin{equation}\label{log}
   \ln\ell(\pi,\theta\mid\boldsymbol{n},\hat\gamma)=\boldsymbol{n}'\ln\boldsymbol\pi^*, 
\end{equation}
where $\boldsymbol{n}$ is the vector with the observed randomized response frequencies $n_{ys}$ corresponding to the elements $\pi^*_{y\mid s}$. If the model includes the parameter $\gamma$, then $\hat\gamma$ is treated as a fixed value.

\section{Methods}\label{correction}

In this section we derive the two methods to correct for random answering. The first one estimates $\gamma$ on the basis of the error rate in the answers to the control question. The second method uses the completion time of the survey as a proxy for random answering, and gives less weight in the log-likelihood to respondents who completed the survey in an unrealistically short time. 

\subsection{Estimation of $\gamma$ on the basis of the control question}\label{sub:gamma}

If all respondents know the correct answer to the control statement, the prevalence of random answering is estimated as $\hat\gamma = 2\cdot{e_c}$, where $e_c$ denotes the observed proportion of errors on the control question. The multiplication by two is necessary because on average half of the random responders will answer the control question correctly by chance. The prevalence estimate of the sensitive attribute is then obtained by plugging $\hat\gamma$ in for $\gamma$ in the random responders model (\ref{mod:random}), or in the model (\ref{mod:onesayer-rr}), that additionally accounts for one-saying.

For the data at hand, however, we expect that some respondents answered the control question incorrectly because they did not know the correct answer (see Section \ref{data}). If we denote the prevalence of these respondents by $\phi$, then $e_c=.5\gamma + \phi$ (see Appendix A for a derivation) is a mixture of random respondents with prevalence $\gamma$ and respondents who think they are not licensed/accredited with prevalence $\phi$. Under the assumption that $\phi$ is independent of doping use, the conclusion is justified that $\phi$ has no effect on the validity of the answers to the doping statement. The problem then reduces to the estimation of $\gamma$.

The procedure to estimate $\gamma$ is as follows. Let $\hat\pi_{in}$ and $\hat\pi_{out}$ respectively denote the doping prevalence estimate of the ECWM  (\ref{mod:adherence}) with the respondents who answered the control question incorrectly in- and excluded from the data. Their exclusion implies that approximately half of the random respondents are eliminated from the data, since the other half answered the control question correctly by chance. If random answering is completely absent we expect that $\hat\pi_{in}\approx\hat\pi_{out}$. If not, the expectation is that $\hat\pi_{in}>\hat\pi_{out}$, because random answering biases the prevalence estimate toward 0.5. Since the exclusion of the incorrect answers to the control question excludes only half of the random responders, the difference between the two estimates $\Delta\hat\pi=\hat\pi_{in}-\hat\pi_{out}$ is due to $.5\gamma$. Given the linearity of the relationship between $\mathbb{E}(\hat\pi)$ and $\gamma$ as depicted in Figure \ref{fig:bias}, the estimate $\hat\pi_{ra}$ that is corrected for random answering is thus given by
\begin{equation}\label{eq:gammahat}
    \hat\pi_{ra}=\hat\pi_{out} - \Delta\hat\pi.
\end{equation}
The estimate of $\gamma$ is then found by fitting the random responders model (\ref{mod:random}) to the data with all respondents included with a fixed value $\hat\gamma$ that yields the prevalence estimate $\hat\pi_{ra}$. The same procedure is followed for the one-sayers model (\ref{mod:one-saying}) by estimating $\hat\pi_{one,~in}$ and $\hat\pi_{one,~out}$ with the one-sayers model, and defining $\Delta\hat{\pi}_{one}=\hat\pi_{one,~in}-\hat\pi_{one,~out}$ and $\hat\pi_{ra+one}=\hat\pi_{one,~out}-\Delta\hat{\pi}_{one}$.

\subsection{Weighting on the basis of completion time}

\begin{figure}[!ht]
    \centering
    \includegraphics[width=.7\textwidth]{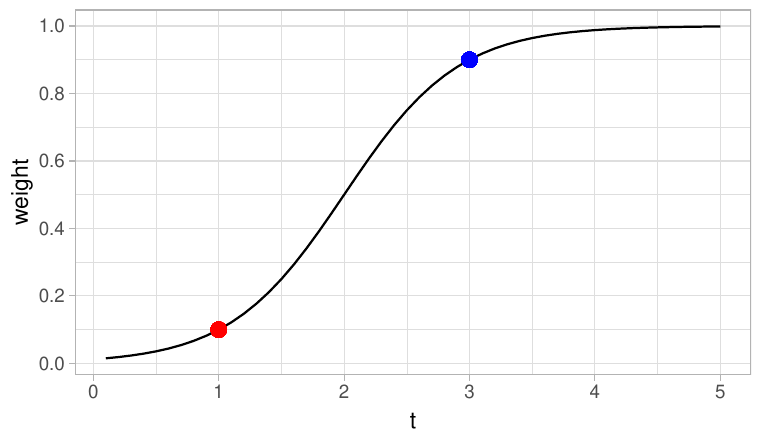}
    \caption{Weight $w$ as a function the survey completion time $t$ for $t_0=1$ and $t_{.5}=3$ minutes.}
    \label{fig:weight}
\end{figure}

Respondents who completed the survey in a very short time are more likely to have skipped the instructions and to have answered the statements randomly, in comparison with respondents who took a longer time to complete the survey. A solution to account for random responding is to exclude fast respondents from the analysis, at the risk of also excluding (fast) respondents who did not answer randomly. Alternatively, we propose maximization of the weighted log-likelihood function for the most general case that includes corrections for the one-saying and random answering based on the control statement
\begin{equation}\label{eq:loglw}
   \ln\ell(\pi,\theta\mid\boldsymbol{y, w, \hat\gamma})=\sum_iw_i\ln\pi^*_i, 
\end{equation}
with $w_i$ denoting the probability that respondent $i$, for $i=1,\dots,n$, is a not random responder. For estimation of these weights we propose the logistic regression model 
\begin{eqnarray}\label{eq:weights}
\mbox{logit}(w_i)=\log\frac{w_i}{1-w_i}=\beta_0 +\beta t_i,
\end{eqnarray}
where $t_i$ denotes the completion time of respondent $i$. The parameters $\beta_0$ and $\beta$ should be chosen such that the weights $w_i$ increase with the completion time. This can be achieved by solving 
\begin{eqnarray}\label{beta}
  \left(\begin{array}{c}
 \beta_0\\
 \beta 
  \end{array}\right)=
\left(\begin{array}{cc}
  1   & t_{0}\\
  1   & t_{.5}
  \end{array}\right)^{-1}
  \left(\begin{array}{c}
  \mbox{logit}(w_0) \\
  \mbox{logit}(w_{.5})
  \end{array}\right),
\end{eqnarray}
where $t_0$ and $t_{.5}$ are respectively the fastest and median completion times observed in the sample, and $w_{0}$ and $w_{.5}$ the corresponding weights. We propose to use $w_0=0.1$, and  $w_{.5}=0.9$, thus giving the fastest respondent a probability of 0.9 of random answering, and a respondent with the median completion a probability of 0.1 of random answering. Figure~\ref{fig:weight} illustrates the weight function for $t_0 = 1$ minute (red point) and $t_{.5} = 3$ minutes (blue point), yielding $\beta_0= -4.39$, and $\beta=2.19$.

\hypertarget{results}{%
\section{Results}\label{results}}

In this section we present the prevalence estimates of doping use for the Surveys A to D of the ECWM (\ref{mod:adherence}), the one-sayers model (\ref{mod:one-saying}), the one-sayers model with correction for the control statement  (\ref{mod:onesayer-rr}), and the weighted version of this model (\ref{eq:loglw}).

Before presenting the parameter estimates of these models, we show in Table \ref{tab:gamma} how the estimates of $\gamma$ and the $\beta$ parameters for the one-sayers model were obtained. The column $2e_c$ shows twice the error rates observed on the control statement, which would have been our estimate of $\gamma$ if all athletes would have known that they were accredited/licensed.  The columns $\hat\pi_{one, ~in}$ and $\hat\pi_{one,~out}$ show the  prevalence estimates of doping use obtained with the one-sayers model, and $\Delta\hat\pi_{one}$ is the difference between the two. As expected in the presence of random answering, these differences are positive. The $\hat\pi_{one+ra}=\hat\pi_{one, ~out}-\Delta\hat\pi_{one}$ are prevalence estimates corrected for random answering. The values of $\hat\gamma$ are obtained by fitting the one-sayers model to the data with all respondents included, with $\hat\gamma$ chosen such that these models yield $\hat\pi_{ra+one}$. The $\hat\gamma$ values are all smaller than $2e_c$, suggesting that part of incorrect answers to the control question are due to ignorance with respect to the accreditation/licensing.

\begin{table*}[!ht]
\caption{Estimation of $\hat\gamma$ and the $\beta$ parameters for the one-sayers model.}
\label{tab:gamma} 
\centering
\begin{footnotesize}
\begin{threeparttable}
\begin{tabular}{p{4em}lccccccccccccccccc}
\toprule
& $2e_c$ & $\hat\pi_{one,~in}$ & $\hat\pi_{one,~out}$ &  $\Delta\hat\pi_{one}$ & $\hat\pi_{ra+one}$ &$\hat\theta$ &$\hat\gamma$ & $t_0$ & $t_{.5}$ & $(\beta_0, \beta)$\\ \hline
\addlinespace[1em]
{A}  & .254 &  .130   & .114  &  .016 & .098 & .106 & .072 &  1.4 & 3.4  & (-5.27, 2.20)\\                           
{B}  & .270 &  .357   & .331  &  .026 & .306 & .115 & .233 &  0.7 & 2.8  & (-3.66, 2.09)\\
{C}  & .394 &  .213   & .194  &  .019 & .176 & .131 & .099 &  0.9 & 2.9  & (-4.17, 2.20)\\
{D}  & .268 &  .129   & .126  &  .003 & .124 & .088 & .013 & 1.5  & 3.5  & (-5.49, 2.20)\\ 
\bottomrule
\end{tabular}
\end{threeparttable}
\end{footnotesize}
\end{table*}

The fastest and median completion times $t_0$ and $t_{.5}$ are then used to compute $\beta_0$ and $\beta$ according to Eq. (\ref{beta}) with $w_0=0.1$ and $w_{.5}=.9$, and these $\beta$'s are in turn used to compute weights $w_i$ for the one-sayers model corrected for the control statement with the weighted log-likelihood function (\ref{eq:loglw}). As in the analysis of these data with the one-sayers model \citep{cruyff2024one}, completion times in excess of 15 minutes are excluded from the data because they are most likely due to a failure to stop the timer, and their inclusion might seriously bias the parameter estimates. The exclusion of these cases reduces the sample sizes of surveys A-D by 6\%, 2\%, 1\% and  5\%, respectively. The fastest completion times in the four surveys range from 0.7 to 1.5 minutes, and the median completion times from 2.9 to 3.5 minutes. The $\beta$ values to obtain the weights $w_0=.1$ and $w_{.5}=.9$ are shown in the last column. We conducted a sensitivity analysis to examine the effects of using different values for $w_0$ and $w_{.5}$ on the prevalence estimates. Appendix B on OSF (\url{https://osf.io/ekcjb/?view_only=e5ad20e51f2c4b4ea3816d5281350782}) summarizes the results for the four surveys of using $w_0\in\{.01, .1, .2\}$ in combination with $w_{.5}\in\{.8, .9,.99\}$, and shows that these effects are relatively small.

\begin{table*}[!ht]
\caption{Uncorrected and corrected doping prevalence estimates $\hat\pi$.}
\label{tab:all} 
\centering
\begin{footnotesize}
\begin{tabular}{p{5em}lcccl}
\toprule
Survey & Model & $\hat\pi$ (95\% CI) & $\%\hat\pi_{ECWM}$ & $\hat\theta$ (95\% CI) & goodness-of-fit\\
\hline
\addlinespace[1em]
A & ECWM            & .166 (.086, .246)  & 100   &                    & $G^2_{1}=4.8, ~ p=.028$\\
  & + one-saying    & .130 (.039, .222)  & 78.3  & .106 (.010, .202)  & $G^2_{0}=0$\\
  & + ra$^*$        & .098 (.000, .220)  &  59.0 & .106 (.010, .202)  & $G^2_{0}=0$\\
  & + weights$^*$   & .055 (.000, .177)   &  33.1 & .111 (.001, .221)  & $G^2_{0}=0$\\
\addlinespace[1em]
B & ECWM            & .374 (.285, .462)  & 100  &                    & $G^2_{1}=4.4, ~ p=.036$\\
  & + one-saying    & .357 (.225, .485)  & 95.4 & .115 (.009, .222)  & $G^2_{0}=0$\\
  & + ra$^*$        & .306 (.181, .431)  & 81.8 & .115 (.009, .222)  & $G^2_{0}=0$\\
  & + weights$^*$   & .280 (.147, .413)  & 74.9 & .119 (.001, .237)  & $G^2_{0}=0$\\
\addlinespace[1em]
C & ECWM            & .251 (.200, .302) & 100  &                     & $G^2_{1}=17.3, ~ p<.001$\\
  & + one-saying    & .213 (.152, .273) & 84.9 & .131 (.070, .192)   & $G^2_{0}=0$\\
  & + ra$^*$        & .176 (.090, .262) & 70.1 & .131 (.070, .192)   & $G^2_{0}=0$\\
  & + weights$^*$   & .161 (.071, .248) & 64.1 & .138 (.070, .207)   & $G^2_{0}=0$\\
 \addlinespace[1em]
D & ECWM            & .161 (.108, .214) & 100  &                     & $G^2_{1}=7.5, ~ p=.006$\\
  & + one-saying    & .129 (.069, .189) & 80.1 & .088 (.025, .151)   & $G^2_{0}=0$\\
  & + ra$^*$        & .124 (.051, .197) & 77.0 & .088 (.025, .151)   & $G^2_{0}=0$\\
  & + weights$^*$   & .107 (.027, .182) & 66.5 & .067 (.000, .139)   & $G^2_{0}=0$\\
\bottomrule
\end{tabular}
\begin{tablenotes}
  \item[] $^*$~~~95\% confidence intervals of $\hat\pi$ obtained with the non-parametric bootstrap.
\end{tablenotes}
\end{footnotesize}
\end{table*}

Table \ref{tab:all} shows the prevalence estimates of of doping of the four models. The goodness-of-fit statistics show that for none of the surveys the ECWM  fits the data, and therefore we started by fitting the one-sayers model. This model is saturated, and therefore yields a zero $G^2$ statistic. The column with $\%\hat\pi_{ECWM}$ shows the percentage reduction in the corrected estimates relative to the uncorrected estimates of the ECWM. For instance, in survey A the corrected estimate for one-saying and random answer is $(.098/.166)*100 =59\%$ of the uncorrected estimate of the ECWM. The correction for one-saying results in a reduction of the uncorrected estimates of approximately 5\% to 20\%. The additional corrections for the control question and weighting  further reduce the prevalence estimates compared to the model without the weights. The model with all corrections yields prevalence estimates that are between 25.1\% and 35.9\% lower than the uncorrected estimates, but for Survey A the difference of 66.9\% is much larger. Here 21.7\% is due to one-saying, so that the further reduction of 45.2\% is due to random answering. 

To account for the uncertainty in the estimate $\hat\gamma$, the 95\% confidence intervals of $\hat\pi_{one+ra}$ and $\hat\pi_{one+ra+w}$ (the  weighted estimate of doping corrected for random answering) are obtained with the non-parametric bootstrap. For $10, 000$ bootstrap samples of the doping statement and the control question $\hat\pi_{one+ra}$ and $\hat\pi_{one+ra+w}$ are estimated in the same way as for the original data, and from these estimates the 95\% confidence intervals are obtained with the percentile method. 



\hypertarget{Discussion}{%
\section{Discussion}\label{discuss}}
 
In the present paper, we proposed two methods to estimate and correct for random answering of the (E)CWM. One method is to estimate the prevalence of random responders by employing a non-sensitive control statement for which both the true and randomized answers are known at the individual level, and the other method is to incorporate individual weights in the log-likelihood function to obtain a weighted estimator of the sensitive attribute using respondents' times to complete the survey/CWM statements. Each method addresses a different aspect of random answering. The former method identifies (half of the) random responders by their incorrect answers to the control statement. This may include inattentive respondents who answer fast, but also respondents who try their best to comprehend the instructions but fail to do so and thus take longer to complete the survey. The latter method identifies random responders stochastically by unrealistically fast completion times. Both methods can be applied independently of each other, but are most effective when applied in conjunction with each other.

The analyses of the data have shown that our corrections for random answer invariably result in substantially lower prevalence estimates. These results are in line with the previous studies \citep{hoglinger2016, hoglinger2018, enzmann2017, schnapp2019,hoglinger2017}, suggesting that higher prevalence estimates of sensitive attributes cannot be interpreted as evidence for a successful control of evasive response bias and random answering provides an alternative explanation for these high estimates. Unfortunately, the significance of these effects cannot be tested, as the corrections for random answering do not affect the goodness-of-fit of the model. For the data at hand, the quality of the estimates may have been be negatively affected by measurement errors in the answers to the control statement due to the ignorance of being accredited/licensed, and in completion times due to failures to stop the timer. These measurement errors unnecessarily complicated the analyses but, as we will show, can be easily avoided by taking the necessary precautions.

If all athletes would have been aware of being licensed/accredited, $\gamma$ could have simply been estimated by twice the error rate $e_c$ on the control statement. Since this was not the case, the complicated method presented in Section \ref{sub:gamma} and Table \ref{tab:gamma} had to be applied to distinguish between random responders and respondents who were unaware of being accredited/licensed. While this method seems to be valid under the assumption that the difference between the estimates with and without the incorrect answers to the control statement are due to random answering, it decreases the reliability of the $\gamma$ estimates. These errors can  be avoided by formulating an unambiguous control statement. An example is to tell respondents that they are shown a randomly selected picture, and ask them if it is the picture of the airplane. The alternative pictures may represent an apple or a house, and since the selected picture is known to the researcher, the respondent's answer to this statement is known. Combined with a quasi-randomized number sequence statement this would leave no room for ambiguity. 

The errors in the completion times measurement also complicated the analyses. To avoid bias in the parameter estimates, it was necessary to remove cases with unrealistically long completion times from the data, resulting in a loss of data. To distinguish between realistic an unrealistic completion times we used the somewhat arbitrary cut-off point of 15 minutes. The
main reason for such long completion times may have been the failure to click the ``Submit'' button, but there are also other explanations. Respondents may have taken a break while filling out the survey, for example by getting a cup of coffee or chatting with an acquaintance. We therefore recommend that the completion times be measured under controlled circumstances where potential distractions are avoided as good as possible. We also recommend that the times to read the instructions and answering the ECWM questions be measured separately, so that occasional outliers due to distractions can be taken into account.

A remarkable finding was the high error rate of 19.7\% on the control question of Survey C, while in the other three surveys it range from 12.7\% to 13.5\%. A substantial part of the surveys were administered in the registration hall where the athletes collected their accreditation pass, so that it seems likely that these athletes should have been aware of being accredited. While it is important to understand the causes of this high error rate, we do not have a conclusive explanation for it as yet. In a future study we will explore potential explanations for it.

To sum up, the validity and reliability of the corrections for random answering presented in this paper are not optimal due to measurement errors in the answers to the control question and in the completion times. In order to avoid such measurement errors, we suggested the use of an improved control question and separate time measurements for reading the instructions and answering the questions. In future surveys we will effectuate these suggestions and if they have the anticipated effects on the measurement errors, the two proposed methods promise to be powerful tools to estimate and correct for random answering.

\newpage
\bibliography{Ref}

\end{document}